\documentstyle[11pt,twoside,paspfoo,epsf]{article}

\begin{document}

\title{Summary of Recent Progress in Understanding HVCs}

\author{Bart P. Wakker}
\affil{Department of Astronomy, University of Wisconsin, Madison}
\author{Hugo van Woerden}
\affil{Kapteyn Institute, Rijks Universiteit Groningen}
\author{Brad K. Gibson}
\affil{Center for Astrophysics \& Space Astronomy, University of Colorado}

\def\kms{km\,s$^{-1}$}
\def\Ha{H$\alpha$}
\def\HI{H{\small I}}
\def\HII{H{\small II}}
\def\CII{C{\small II}}
\def\CIV{C{\small IV}}
\def\NI{N{\small I}}
\def\NII{N{\small II}}
\def\OIII{O{\small III}}
\def\OVI{O{\small VI}}
\def\MgII{Mg{\small II}}
\def\PII{P{\small II}}
\def\PV{P{\small V}}
\def\SI{S{\small I}}
\def\SII{S{\small II}}
\def\SIV{S{\small IV}}
\def\SVI{S{\small VI}}
\def\ArI{Ar{\small I}}
\def\CaII{Ca{\small II}}
\def\SiII{Si{\small II}}
\def\SiIII{Si{\small III}}
\def\ZnII{Zn{\small II}}

\section{Introduction}
\par The study of high-velocity clouds has progressed much since the appearance
of the review article by Wakker \& van Woerden (1997), less than two years ago.
Much of this progress is described in these workshop proceedings. Here we update
the review article, summarizing the topics discussed at the workshop as well as
covering the recent literature. We follow the outline of the review, describing
\HI\ properties in Sect.~2 and 3, interactions between HVCs and other gas in
Sect.~4, observations at wavelengths other than 21-cm in Sects.~5 and 6,
absorption-line studies of metallicities and distances in Sects.~7 and 8,
extra-galactic HVCs in Sect.~9, the Magellanic Stream in Sect.~10, and a
discussion of HVC origins in Sect.~11.
\par Key contributions of the past two years include (a) the first determination
of a distance {\it bracket} for an HVC: $d$=4--10\,kpc for complex~A (van
Woerden et al.\ 1999); (b) the first determinations of a true metallicity:
$\sim$0.25\,Z$_\odot$ for HVC\,287+22+240 (Lu et al.\ 1998), and
$\sim$0.1\,Z$_\odot$ for complex~C (Wakker et al.\ 1999); (c) the recognition of
a leading counterpart to the Magellanic Stream (Gardiner \& Noguchi 1996, Lu et
al.\ 1998, Putman et al.\ 1998), indicative of a tidal origin; (d) the quickly
growing number of optical emission line observations (primarily \Ha) of HVCs,
led by Tufte et al.\ (1998) and Bland-Hawthorn et al.\ (1998); and (e) the
proposal by Blitz et al.\ (1998) that HVCs are intergalactic remnants of the
formation of the Local Group.

\section{Distribution and structure of HVCs}
\par While all-sky \HI\ surveys for HVCs have existed since the late 1980s, a
major step forward in ensuring uniformity in spatial and velocity resolution can
be expected after the release of the Villa Elisa Southern Sky Survey (Arnal et
al.\ 1996), a direct analog to the Leiden-Dwingeloo Survey (LDS) (Hartmann \&
Burton 1997) in the north. Together, these will cover the sky at 30\arcmin\ and
1\,\kms\ resolution. For clouds between declinations of $-$90\deg\ and 0\deg,
higher angular resolution (16\arcmin\ on a sub-Nyquist sampled spatial grid,
with 13\,\kms\ velocity resolution) will ultimately be available when the Parkes
Multibeam Survey (Staveley-Smith 1997) completes its scans of the sky.
\par A crucial first step in the analysis of the combined surveys will be the
creation of a catalogue for all of the anomalous-velocity gas. Blitz reported
that this work is in progress. The catalogue will contain all of the gas at
velocities incompatible with a simple model of differential galactic rotation,
and for the first time will contain reliable information on linewidths, making
it possible to address some of the questions raised in the review of Wakker \&
van Woerden (1997, Sects.~2.3--2.7). Cross-correlation with the Parkes Multibeam
HVC Survey will shed light on any residual compact, isolated HVCs missed by the
somewhat coarser LDS and VESSS.
\par Using these new surveys, the questions of HVC sky coverage and the
distributions of brightness temperature, cloud flux and cloud area can be
addressed again. A proper discussion of the selection effects for low brightness
temperatures and small clouds should allow to decide whether the apparent
turnovers at low values in these distributions are real or not. Blitz et al.\
(1998) argue they are real and thus conclude that HVCs have a typical size of
1.5 degrees, typical distance 1\,Mpc and a typical \HI\ mass of
3$\times$$10^7$\,M$_\odot$. Wakker \& van Woerden (1997, Sect.~2.5) point out
that the turnover in the N(\HI) distribution found by Giovanelli (1980) appears
to shift to lower N(\HI) for the Hulsbosch \& Wakker (1988) data. Including the
limited, but more sensitive results of Murphy et al.\ (1995) makes the turnover
move to even lower values of N(\HI), indicating that selection effects are
probably important.

\section{Small-scale structure}
\subsection{High-resolution observations}
\par Putman et al.\ (priv.\ commm.) reported the initiation of a long-term
high-resolution ($\sim$1\arcmin) study of a number of southern HVCs with the
Australia Telescope Compact Array (ATCA) at Narrabri. Wakker et al.\ (these
proceedings) report on the ATCA observations of HVC\,287+22+240. The ATCA has
also mapped a HVC serendipitously: Oosterloo et al.\ (1996) found a small \HI\
cloud in the field of the newly discovered Tucana Local Group dwarf; this cloud
(HVC323$-$48+135) is probably associated with the Magellanic Stream. Several
Anti-Center HVCs were mapped at 3\arcmin\ resolution by Tamanaha (1995, 1997). A
few northern HVCs were mapped with the Westerbork Synthesis Radio Telescope
(WSRT), including a field in complex~L, HVC\,100$-$7+100 (Stoppelenburg et al.\
1998), and the Mark\,290 field (see Sect.~7).
\par The primary use of such data often is the derivation of accurate \HI\
column densities in the direction of extragalactic and stellar probes, in order
to determine accurate metallicities and to establish significant limits on
non-detections, allowing the derivation of lower limits to distances. A second
important use is the study of fine structure in the ISM.
\par The importance of high angular resolution for abundance studies was clearly
shown by Lu et al.\ (1998) for the case of HVC287+22+240. In the direction of
the background Seyfert galaxy NGC\,3783 the value of N(\HI) derived from a
21\arcmin\ beam (Green Bank 140-ft) is a factor 1.5 higher than the value
derived from a 1\arcmin\ beam (ATCA), changing the derived sulphur abundance
from 0.15 to 0.25 times solar. The latter value indicates Magellanic-Cloud like
metallicities. Since this cloud is in a part of the sky where the tidal model
for the Magellanic Stream of Gardiner \& Noguchi (1996) predicts Magellanic gas
to be present, the abundance result supports a relation to the Magellanic
Stream. Using the uncorrected value would have given more weight to the idea
that HVC287+22+240 could be an independent intergalactic object.

\subsection{Analyses of high-resolution maps}
\par High-resolution maps of high-velocity clouds provide a convenient
laboratory to study fine structure in the ISM, as the HVCs are relatively
isolated both in velocity and position, and thus there is little confusion with
foreground and background emission. The simplest analysis is to look at column
density contrasts. Wakker \& Schwarz (1991) and Wakker et al.\ (1999) find that
the structure is hierarchical, consisting of cores embedded in a smoother
background. At the cores, the column density contrast can be as high as a factor
5 over a few arcminutes. Off the cores factors of 2 on arcminute scales are
common. A slightly more complex analysis was done by Vogelaar \& Wakker (1994),
who find that the structure can be described by a fractal with dimension 1.4,
similar to the values found for molecular clouds (Falgarone et al.\ 1991). Other
methods of analysis based on the Fourier plane or Minkowski functionals have
been suggested, but not yet implemented.
\par A new use for high-angular-resolution maps of HVCs was suggested by
Ivezi\'c \& Christodoulou (1997) and Christodoulou et al.\ (1997). The former
paper looked for Young Stellar Objects (YSOs) associated with the small \HI\
cores in cloud MI and the core of complex H that were observed by Wakker \&
Schwarz (1991). Ivezi\'c \& Christodoulou (1997) point out the coincidence of
the only YSO in the MI field with the brightest \HI\ knot (MI core 4 in Wakker
\& Schwarz 1991). Three further YSOs are found in the core of complex~H.
Christodoulou et al.\ (1997) analyze the likelihood of forming stars inside HVCs
by means of the collision of cloudlets and find that at most a few stars should
form over several Gyr, consistent with the observed numbers. However, no
estimate was made of the probability that these coincidences are due to chance
superpositions.

\subsection{Observational determination of pressure}
\par Theoretically, the gas pressure is one of the most important concepts.
Wolfire et al.\ (1995) show how the assumption that the clouds are confined by
an external pressure leads to a two-phase core-envelope structure. However, it
is not trivial to measure the cloud pressure observationally. A possible way of
deriving pressure using high-resolution \HI\ data will be described by Wakker et
al.\ (1999). This method is summarized here. First, use the maps to derive the
volume density for each core (see details below); second, fit one or more
gaussians to the profile at each pixel and convert the FWHM linewidth(s) $W$ to
a temperature: $T(x,y)$=21.8\,$W^2(x,y)$; third, for each core make a histogram
of the product $P(x,y)$ = $n(x,y)$\,$T(x,y)$. The derived pressure is then given
by the median of this distribution. To derive the volume density $n(x,y)$, start
by defining a box around each core, then calculate the number of pixels with
N(\HI) above 0.5 times the maximum value in that core, and convert to a core
diameter ($\alpha$ = $\sqrt{{4\over\pi}\,N_{\rm pix}}$ $\times$ gridspacing).
Assuming that the line-of-sight density profile is gaussian, then $n(x,y)$ =
N(\HI)$(x,y)$ / [1.064\,$\alpha$\,3.08$\times$$10^{21}$\,D(kpc)].
\par In practice, one finds that the derived radius increases with the square
root of the angular resolution ($\theta$), while the peak column density
decreases as $\theta^{1/2}$. Thus, the pressure derived in this manner varies as
1/$\theta$. Further, within a single core the pressure histograms tend to be
wide (rms dispersion/median $\sim$0.3 in the best cases). The interpretation of
this result is unclear. It may imply that the cores are unresolved; or the
pressure may vary throughout the core, so that at different resolutions
different averages are found; or the density structure may be more complex than
that of a gaussian. At the highest resolution (1\arcmin) the values that are
found for the cores are of order 20000/D(kpc)\,K\,cm$^{-3}$. This is a
combination of the thermal and ``turbulent'' pressure. Assuming a temperature of
50\,K (as measured for complex~H by Wakker et al.\ 1991), the thermal pressure
in the core (which is the number to be compared to the Wolfire et al.\ (1995)
models) would be 1500/$\theta$(arcmin)/D(kpc)\,K\,cm$^{-3}$. If one does the
same exercise for the large-scale structure of the HVCs, pressures of order
5000/D(kpc)\,K\,cm$^{-3}$ are derived for the cloud envelopes. These are clearly
lower than the pressures found from the highest-resolution data. A more detailed
analysis of observational selection effects and biases is required to understand
the problems mentioned above.

\section{Interactions between HVCs and other gas}
\par Many papers claim evidence for interactions between HVCs and gas in the
galactic disk (see Sect.~2.9 in Wakker \& van Woerden 1997 and below). In almost
all cases this is based on a comparison of the morphology of the HVC with that
of lower-velocity gas, combined with a consistency argument concerning the
energetics. However, hard evidence will require knowledge of the distances to
all \HI\ features in the regions studied.
\par On the theoretical side, many models of impacting clouds exist (e.g.
Comeron \& Torra 1992, 1994), whose phenomenology can be compared to the
observations. A different approach was taken by Benjamin \& Danly (1997), who
studied the effects of drag. They argued that many of the clouds falling towards
the Galactic plane in the solar neighborhood are traveling at or near terminal
velocity, which is determined by balancing the drag of the intercloud medium
with the gravitational attraction of the Galaxy. This model is expected to work
best for the closer IVCs (see contribution by Gladders et al.), where gas
density and thus drag are highest. For the more distant high velocity clouds,
the effects of drag on the dynamics of infall should be considered, as it will
affect how quickly the orbits of extra-galactic infalling objects, such as the
Magellanic Stream, will decay.
\par Observationally, Tamanaha (1995, 1996, 1997) used Arecibo to map (at
3\arcmin\ resolution) the Anti-Center Shell (an arc at ($l$,$b$,v$_{\rm LSR}$)=
(180,0,$-$70) found by Kulkarni et al.\ 1985), ACI (($l$,$b$,v$_{\rm LSR}$)=
(182,$-$11,$-$180)), ACII (($l$,$b$,v$_{\rm LSR}$)= (185,$-$15,$-$200)), and the
Cohen Stream (($l$,$b$,v$_{\rm LSR}$)= (165,$-$46,$-$260)). In this series of
papers evidence is presented that the HVCs and IVCs in the anti-center are
falling onto the disk. In particular, the shape of the apparent cavity in the
disk gas is similar to that expected from an oblique impact in the models of
Comeron \& Torra (1992).
\par The possibility that infalling HVCs triggered star formation in the Orion
region was discussed at the workshop by Lepine (see also Lepine \& Duvert 1994),
who showed suggestive maps of the distribution of the OB associations, relative
to that expected from an infalling cloud sheared by differential galactic
rotation. This is the same area of sky studied by Tamanaha, although the
suggested direction of infall is opposite. This may not be a problem, as the
ACI, ACII, etc clouds may be more distant than the Orion star-forming region.
\par Finally, Morras et al.\ (1998) present an Effelsberg map of complex~H, and
argue that the edge of the galactic \HI\ distribution has a hole in this
direction, caused by the infall of complex~H. The evidence is weak, however.

\section{Optical emission lines}
Pre-1996 attempts at measuring ionized hydrogen at high velocity, through
measurement of its \Ha\ emission, were summarized by Wakker \& van Woerden
(1997, Sect.~3.2). Two processes are capable of producing \Ha\ emission;
discriminating between these two can lead to a fuller appreciation of the
environment in which the HVCs reside. If photoionization dominates, the \Ha\
intensity directly reflects the Lyman continuum flux incident upon the cloud; if
collisional ionization dominates, the \Ha\ intensity would be a reflection of
the ambient density and the relative velocities between the cloud and the medium
through which it moves. Unfortunately, determining which is dominant is
generally not possible based upon \Ha\ alone, but requires emission line ratios
(e.g. [\NII],[\OIII],[\SII]/\Ha).
\par The past year has seen rapid progress being made in this field, driven
observationally by two instruments --- the Wisconsin H$\alpha$ Mapper (WHAM)
(Tufte et al.\ 1998) and TAURUS-2 (Bland-Hawthorn et al.\ 1998). An important
attribute of WHAM is its association with a dedicated 0.6\,m telescope. The
aperture advantage of TAURUS-2 on the 4\,m AAT and WHT is tempered by the
oversubscription rates of both telescopes. For HVC work, a major advantage
enjoyed by WHAM is its velocity resolution ($\sim$12\,\kms, vs 45\,\kms\ for
TAURUS-2), which is well suited for resolving the typically $\sim$20\,\kms\ \Ha\
linewidths. On the other hand, TAURUS-2 has a larger bandpass ($\sim$50\,\AA\ vs
$\sim$5\,\AA), allowing simultaneous observations of several emission lines. The
large field-of-view of WHAM ($\sim$1\deg) is superior for mapping the large
complexes, while the 5\arcmin\ field-of-view of TAURUS-2 is superior for compact
HVCs and situations for which emission measures close to a given line-of-sight
are desired.
\par Tufte et al.\ (1998) report H$\alpha$ detections toward complexes~M, C and
A. For complex~M, I(\Ha) ranges from 60 to 200\,mR, for complex~A it is 85\,mR,
and for the single complex~C pointing, 130\,mR. The quoted measurement errors
are typically $\sim$10--20\%. In two complex~M directions [\SII]$\lambda$6716
was also observed, in an attempt at constraining the ionization mechanism. The
[\SII]/\Ha\ ratios are 0.64$\pm$0.14 for the direction with the highest N(\HI),
and $<$0.11 one degree away, where there is little \HI. The former is suggestive
of photoionization, while the latter is consistent with a shock-induced origin.
Clearly, there exist significant variations in the ionization conditions within
Complex~M.
\par WHAM was also employed by Wakker et al.\ (these proceedings) along the
complex~C sightline sampled by the background Seyfert galaxy Mark\,290. An
I(\Ha) of 190\,mR was found, as well as a 3$\sigma$ upper limit of 20\,mR on
[\SII]$\lambda$6716. This is the first study to combine the constraints imposed
by (i) non-depleted elemental abundances, (ii) 21-cm synthesis column densities,
(iii) emission line intensities, and (iv) stellar probe distance limits,
providing unique insight into the physical conditions (e.g. ionization fraction,
temperature, thermal pressure) at play in the Galactic halo.
\par Parallel to the WHAM Team's efforts, the TAURUS-2 Team presented their
first results (Bland-Hawthorn et al.\ 1998). They detected \Ha\ (270\,mR) and
[\NII]$\lambda$6548 (127\,mR) emission lines from HVC\#360 in complex~GP (Wakker
\& van Woerden (1991), also known as the Smith Cloud. This gives a ratio of
0.47, which is enhanced by a factor of two relative to the Reynolds layer; the
enhancement is close to a factor of four at the core of the line. A 3$\sigma$
upper limit to the [\OIII] line intensity of 120\,mR was also found. Arguments
in support of photoionization as the dominant ionization mechanism were
presented.
\par The newly released model of Bland-Hawthorn \& Maloney (1998) of the
Galactic Halo ionizing radiation field allows to predict HVC distances, under
the assumption that the optical emission lines from HVCs arise due to ionization
from photons leaking from the Galactic Disk (with some assumptions about the
geometry). In that case the number of emitted \Ha\ photons is proportional to
the number of incoming ionizing photons. Bland-Hawthorn et al.\ (1998) apply
this to the detections of HVC\#360 and derive an implied distance of
$\sim$26\,kpc. They suggest that at this distance an association with the
Sagittarius dwarf is possible. Further testing of the Bland-Hawthorn \& Maloney
radiation field model will be a natural byproduct of both the WHAM and TAURUS-2
surveys.
\par Bregman (these proceedings) points out that Weiner \& Williams (1996) also
report directions in the Magellanic Stream where no \Ha\ is detected, although
\HI\ emission is seen in these directions (at low angular resolution). They used
these non-detections to argue that the \Ha\ emission originates from
ram-pressure heating by the movement of the clouds through a tenuous halo. If it
can be shown that there are no holes in the \HI\ (using higher-angular
resolution data such as those from the Parkes Multibeam survey), these
non-detections imply that the ionizing radiation field reaching the Stream is
much lower than expected. This requires an extra optical depth of 1.2 near
14\,eV, or an halo \HI\ column density of 3$\times$$10^{17}$\,cm$^{-2}$. Then
the \Ha\ intensity can no longer be used to estimate cloud distances.

\section{A connection between HVCs and energetic radiation?}
\par Blom et al.\ (1997) suggest that a feature in the distribution of
0.75--3\,MeV $\gamma$-ray radiation may be generated by the interaction of HVCs
with disk gas. These authors find a $\gamma$-ray emission enhancement in the
Ursa Major Window (Lockman et al.\ 1986) ($l$,$b$$\sim$150\deg,55\deg), whose
edges are lined by the HVC complexes A, M and C. However, it is unclear whether
the Ursa Major Window is associated in any way with the HVCs, also leaving open
the association between HVCs and $\gamma$-rays.
\par At the workshop, Mebold reported on the work of Kerp et al.\ (1998), who
analyzed ROSAT data for four $\sim$30\deg x30\deg\ fields containing HVCs (the
low- and high-latitude ends of complex~C, the complex~WA region and a region
near the GCN complex). They show that the large-scale features of the X-ray maps
are well explained by a simple model of foreground emission associated with the
Local Hot Bubble (LHB, Cox \& Reynolds 1987), combined with background emission
associated with a Galactic halo, which is absorbed by all or most of the \HI\
column density; within each field the foreground and background intensities are
assumed constant. Kerp et al.\ (1998) point out excess emission in the region
$l$=100\deg--130\deg, $b$=50\deg, which is near the main line of complex~C
cores. This excess X-ray emission may originate in an interaction between
complex~C and gas in the lower Galactic halo. If so, Zimmer et al.\ (1997)
provide an explanation for the X-ray emission in terms of the dissipation of
magnetic fields in the Galactic halo. However, a connection between the X-ray
excess and intermediate-velocity gas or small-scale structure in the Galactic
halo cannot be completely ruled out.
\par This X-ray excess was originally interpreted as due to variations in the
shape of the Local Hot Bubble (Cox \& Reynolds 1987). A major part of the
argument for a local origin is based on a comparison of the expected X-ray
optical depths in two energy bands (B-band, 75--200\,eV and C-band,
100--300\,eV) of pre-ROSAT observations (McCammon \& Sanders 1990). The
low-velocity \HI\ (presumably at $z$$<$500\,pc) has a column density of
$\sim$$10^{20}$\,cm$^{-2}$. Then the optical depth in B-band is $<$1, while that
in C-band is $>$5. Yet, the B-band and C-band images look similar, which is hard
to explain if the enhancement originates at $z$$>$1\,kpc. ROSAT did not have the
spectral resolution to address this problem.

\section{Metallicities}
\par Wakker \& van Woerden (1997, Sect.~4.3) discussed the measurement of \SII\
absorption associated with HVC287+22+240 in the spectrum of the Seyfert galaxy
NGC\,3783. Sulphur is one of a few elements that are not depleted onto dust in
the ISM, and thus can provide an absolute metallicity measurement. Lu et al.\
(1994) derived an abundance of 0.15 times solar, but this was based on a
low-resolution \HI\ observation. Wakker et al.\ (these proceedings) observed the
HVC at 1\arcmin\ resolution, using which a sulphur abundance of 0.25$\pm$0.07
times solar was derived by Lu et al.\ (1998). Elsewhere in these proceedings,
Wakker et al.\ report the second detection of sulphur in a HVC. They find
[S/H]=0.094$\pm$0.019$^{+0.022}_{-0.018}$ times solar for complex~C, using
Mark\,290 as the background probe. This is a very low value, and Wakker et al.\
argue that complex~C represents the infall of low-metallicity gas onto the
Galaxy. So far, an absolute metallicity has been found only for these two HVCs,
and both results are much lower than the solar value. While suggestive, it is
premature to conclude that all HVCs therefore have strongly sub-solar
metallicities.
\par For instance, the [\SII]/\Ha\ ratio of 0.64 measured for cloud MI by Tufte
et al.\ (1998) can be used to argue that the metallicity of MI is near solar.
Assuming solar abundance a temperature of $\sim$7000\,K is required, within the
range of expected values (6000--8000\,K; Reynolds 1985). Pressure equilibrium
with a hot halo (for which Wolfire et al.\ 1995 give P(z)) would occur at a
distance of 3\,kpc. If the abundance were 0.1 or 0.5 solar, then T=20,000\,K
(very unlikely) or 9000\,K (unlikely but possible), respectively, would be
required to get the observed level of [\SII] emission.
\par Sahu \& Blades (1997) and Sahu (1998) present low-resolution (120\,\kms)
data on \SiII, \SiIII\ and \NI\ for HVC\,487, probed by the starburst galaxy
NGC\,1705. This cloud is probably a shred of the Magellanic Stream.
Unfortunately, the low resolution and the fact that the lines are saturated do
not allow measurements of the metallicity.
\par Much progress on HVC metallicities, dust depletion patterns and ionization
structure is expected from the Far Ultraviolet Spectroscopic Explorer (FUSE),
set to be launched in May 1999. The capabilities of FUSE for HVC metallicity
work were described at the workshop by Sembach. This satellite can detect lines
from the dominant ionization stage of several undepleted elements (O, N, P, Ar),
lines of many depleted elements (C, Mg, Al, Si, Cl, Cr, Mn, Fe, Ni), lines of
different ionization stages of the same element (especially C, N, P, S and Fe),
many molecular lines (H$_2$, HD, CO), as well as lines originating in hot gas
(\OVI, \PV, \SIV, \SVI). Long integrations are planned for probes of the
Magellanic Stream, complex~A, complex~M, complex~C (8 probes), HVC287+22+240,
and of highly-ionized HVCs. Other HVCs may also be observed, if the candidate
probes prove to be sufficiently bright.

\section{Distances}
\par Work on HVC distances is proceeding at an accelerated rate. The first
distance bracket has now been found, for complex~A (see van Woerden's article in
these proceedings). \CaII\ absorption in the spectrum of the RR\,Lyrae star
AD\,UMa ($l$,$b$=160\deg,43\deg) sets an upper limit of 10$\pm$1\,kpc. A lower
distance limit of 4$\pm$1\,kpc is provided by the non-detection of \MgII\
absorption toward PG\,0859+593 ($l$,$b$=157\deg,40\deg) (Wakker et al.\ 1996).
This lower limit may have been confirmed by Ryans et al.\ (1997b), using
PG\,0832+676. They give 4.6\,kpc for its distance, although earlier estimates
varied from 1.6 to 18\,kpc (Schwarz et al.\ 1995, Brown et al.\ 1989). Using
N(\HI) derived from a combination of Effelsberg and Westerbork data (Schwarz et
al.\ 1995), and the Ca$^+$ abundance for complex~A (Schwarz et al.\ 1995), the
expected value of N(\CaII) is 5$\times$$10^{11}$\,cm$^{-2}$, which would have
yielded a 10$\sigma$ detection.
\par For two other HVCs an upper distance limit has been set. Clouds MII/MIII
have been shown to be less distant than 4.0\,kpc (Danly et al.\ 1993, Keenan et
al.\ 1995), while a tiny HVC in the direction of 4\,Lac is less distant than
1.2\,kpc (Bates et al.\ 1990). No lower distance limits are known for these
clouds.
\par The case of MII/MIII in particular shows the difficulty of interpreting
non-detections, especially in sightlines with faint \HI. An upper distance limit
of $\sim$4\,kpc is set from the detection of absorption toward BD+38\,2182, a
lower limit of 1.7\,kpc was inferred from the absence of absorption toward
HD\,93521, 25\arcmin\ away. However, as summarized by Wakker \& van Woerden
(1997, Sect.~4.4), and discussed more fully by Ryans et al.\ (1997a), with
sufficiently high angular resolution no associated \HI\ emission is found in the
direction of HD\,93521. Thus a lower distance limit can no longer be inferred.
Danly (priv.\ comm.) points out the possibility of ionized hydrogen in the
HD\,93521 sightline; the absence of the strong UV absorption lines would then
still imply a lower distance limit.
\par The third upper distance limit known is for HVC100$-$7+100, which is seen
in absorption toward 4\,Lac (distance 1.2\,kpc; Bates et al.\ 1990).
Stoppelenburg et al.\ (1998) map this HVC at 1\farcm8x2\farcm3 resolution and
show that the cloud is within the Galactic disk ($\vert$$z$$\vert$$<$150\,pc)
and has a mass of only 0.6\,(D/1\,kpc)$^2$\,M$_\odot$.
\par Some progress has been made for HVC complex~C, for which van Woerden et
al.\ (these proceedings) increase the lower distance limit from 2.5 to 5\,kpc.
Finally, Tamanaha (1996) determined lower limits to the distances of AC0
(160\,pc), AC\,I (650\,pc) and the Cohen Stream (=HVC168$-$43$-$260; 350\,pc).
\par At the workshop, distances were also reported for a number of
Intermediate-Velocity Clouds (IVCs) and high-latitude molecular clouds, which
may or may not be inextricably linked with the HVCs. Gladders et al.\ (1998)
report that the distance of the Draco molecular cloud (Herbstmeier et al.\ 1996
and references therein) at ($l$,$b$,v$_{\rm LSR}$)= (90,+39,$-$25) is between
450 and 650\,pc. This is the same object for which Burrows \& Mendenhall (1991)
found a clear X-ray shadow. The new distance determination thus sets a strict
lower limit of 300\,pc to the z-height of part of the X-ray emission.
\par Further progress in determining HVC and IVC distances will be made by a
coordinated effort that was initiated at the workshop to a) find suitable probe
stars in HVC fields, b) obtain intermediate-resolution spectroscopy for
classification and c) do high-resolution spectroscopy at any of the new 8-m
class telescopes coming on-line in the next five years. Beers (this volume) gave
an overview of the possibilities and problems. In most of the southern sky
finding suitable probes may become relatively easy. The Parkes Multibeam Survey
(Staveley-Smith 1997) will provide high-quality HVC data, while the Hamburg ESO
survey (Wisotzki et al.\ 1996) covers 10000 square degrees at $\delta$$<$0\deg\
down to 16.5 magnitude.

\section{HVCs associated with other galaxies}
\par Observations of extra-galactic HVC-analogues have provided more evidence
for a connection between high- (and/or intermediate-) velocity gas and star
formation intensity, probably via a galactic-fountain-type phenomenon. Schulman
et al.\ (1997a) use the VLA to show that UGC\,12732, a face-on galaxy with low
star formation rate, does not have high-velocity \HI\ gas. On the other hand,
two face-on galaxies with high star formation rate do contain such high-velocity
gas: NGC\,5668 (Schulman et al.\ 1996) and NGC\,1300 (Schulman et al.\ 1997b).
The edge-on galaxy NGC\,891 was shown by Swaters et al.\ (1997) to have \HI\ up
to 5\,kpc. These authors also find that the distribution of high-z \HI\
correlates with that of \Ha\ and the radio-continuum.

\section{The Magellanic Stream}
\par While {\it technically} just one of the 17 HVC complexes defined by Wakker
\& van Woerden (1991), the Magellanic Stream occupies a rather special niche on
the HVC family tree. Beyond the obvious (it being perhaps the single most
striking structure in the \HI\ sky, cutting a swath $>$100\deg\ along a Great
Circle through the South Galactic Pole), a large part of its special nature can
be traced to the fact that it is the only HVC of which we know the source: the
Magellanic System.
\par Since its discovery (Mathewson et al.\ 1974), the debate on the origin of
the Stream has focused on the mechanism whereby this gas attained its
present-day distribution. In the mid-1990s two alternative models seemed
feasible: ram-pressure stripping due to an extended gaseous Galactic halo (Moore
\& Davis 1994), or tidal disruption of the Magellanic System due to its
interaction with the Galaxy (Gardiner \& Noguchi 1996). The strongest
discriminants between these two alternatives are that the tidal model predicts
the existence of a leading counter-feature to the trailing Stream, as well as
the presence of stars in the Stream, while the ram-pressure model predicts that
such features are absent. The non-detection of a stellar Stream as well as the
apparent absence of a leading arm made the ram-pressure model seem likely.
\par However, this situation has now changed. At the workshop, Gardiner reported
on the dissertation of Yoshizawa (1998). Building upon the models of Gardiner \&
Noguchi (1996), he incorporated gas dynamics into the existing n-body framework
(embedding initially compact stellar disks in more extended diffuse gaseous
halos). This demonstrates that stars will not be drawn out along the Stream in
appreciable numbers. Some stellar tidal debris is expected, but it is generally
restricted to a $\sim$10\deg--15\deg\ region surrounding the LMC and appears
clump-like, or in two-to-three dispersed streams. Observationally, Ostheimer et
al.\ (1997) and Majewski et al.\ (1999) identified giant stars down to V$\sim$19
in an annulus around the Magellanic Clouds, and obtained luminosity
classifications and radial velocities with follow-up spectroscopy. These
observers tentatively report an excess of giants at distances expected for tidal
debris from the Clouds. It is intriguing that observed excesses appear in the
regions where the simulations of Yoshizawa (see Figure 1 of Gardiner in these
proceedings) predict that stellar tidal debris should appear.
\par Two further observational pieces of evidence appeared during the past year,
offering definitive proof for the tidal origin of the Magellanic Stream. First,
for HVC\,287+22+240 an accurate absolute metallicity of $\sim$0.25\,Z$_\odot$
was found (Lu et al.\ 1998; see also the paper by Wakker et al.\ in these
proceedings), which is similar to the sulphur abundance of the Magellanic
Clouds. This result is especially noteworthy, because HVC\,287+22+240 lies
spatially and kinematically in a region where the Gardiner \& Noguchi (1996) and
Yoshizawa (1998) models predict gaseous tidal debris to reside.
\par Second, using the first data from the \HI\ Parkes All-Sky Survey (HIPASS),
Putman et al.\ (1998) concluded that what appeared to be disconnected and
discrete HVCs in the region between the Magellanic Clouds and the Galactic
Plane, was in fact a continuous feature. These HVCs are connected by a tendril
of \HI\ emanating from the SMC/Bridge region and extend continuously (both
spatially and kinematically) at least as far as the Galactic Plane ($>$25\deg\
away). Gardiner presented a new grid of pure n-body models incorporating drag
forces (based loosely upon the older Gardiner \& Noguchi 1996 models). These
nicely match this feature. However, this version of the model cannot
simultaneously match the details of the Putman et al.\ \HI\ feature and retain
the \HI\ debris in the region near HVC\,287+22+240; better models are obviously
needed.
\par Both the Lu et al.\ (1998) metallicity determination and the Putman et al.\
(1998) leading \HI\ feature are consistent with the tidal models of Gardiner and
collaborators, but not with ram-pressure models. The problem of the absence of a
stellar stream was solved by Yoshizawa (1998). Thus, we can conclude that in the
past year the tidal origin of the Magellanic Stream has been established beyond
reasonable doubt.

\section{Models}
\subsection{General comments}
\par The most notable recent development concerning the origin of HVCs has been
the revival of the idea that some (or all) of them are extra-galactic. This has
been most clearly expressed in the proposal by Blitz et al.\ (1998) that HVCs
are remnants of the formation of the Local Group. Mallouris et al.\ (1998 and
these proceedings) propose a connection with Ly$\alpha$ absorbers seen in the
spectra of background QSOs; in this case the HVCs would still be in the sphere
of influence of the Galaxy, rather than in the Local Group at large, similar to
the model originally proposed by Oort (1970).
\par The idea of extra-galactic origins gets strong observational support from
the metallicity of $\sim$0.1 times solar derived for complex~C (see Sect.~7
above), and also by the ionization properties found for complex~GCN (Sembach et
al.\ 1998). Toward Mark\,509 and PKS\,2155$-$304 they find strong \CIV\
absorption, but no detectable \CII\ or \SiII\ absorption, nor \HI\ emission. A
photo-ionization model yields low density ($<$10$^{-4}$\,cm$^{-3}$) and pressure
($\sim$2\,K\,cm$^{-3}$), favoring a location in the Local Group or very distant
halo (the semi-empirical formula for the halo pressure given by Wolfire et al.\
(1995) gives 2\,K\,cm$^{-3}$ at $z$$\sim$60\,kpc). \par However, it would be
premature to conclude that {\it all} HVCs must be extra-galactic. The 3-D
hydrodynamical models of de Avillez (these proceedings) may allow to set limits
on which HVCs can be accommodated by the Galactic Fountain and which cannot. We
now discuss these developments in more detail.

\subsection{HVCs as Local Group objects}
\par Blitz et al.\ (1998) propose that HVCs are spread across the Local Group,
concentrating in a filament running through M\,31 and the Galaxy. They conclude
that a typical HVC has an \HI\ mass of 3$\times$$10^7$\,M$_\odot$ and a diameter
of 28 kpc, that the ratio of \HI\ mass to total mass is 0.15, and that the
distance scale of the ensemble is about 1\,Mpc. Previously, this possibility was
not considered because of the following arguments: a) the predicted kinematics
are incompatible with the data; b) the typical distance at which the clouds are
gravitationally stable if all the mass is in the form of neutral hydrogen and
helium is 1--75\,Mpc; c) the presence of two-phase structure requires a
substantial external pressure; d) there are small velocity gradients (a few
\kms) over what would be large linear distances (50\,kpc); e) no analogues in
other galaxy groups have been found.
\par Blitz et al.\ try to answer these objections. Against a) they show that the
sky and velocity distributions can be explained from a simple model of Local
Group formation. They solve b) by pointing out that 90\% of the matter may be
dark, as it is in the outer parts of galaxies; this reduces the stability
distances to 0.1--5\,Mpc. A large ionized fraction can further reduce these. 
Objection c) is answered by suggesting that the dark matter in the cores cools,
settles and the gas becomes self-shielded from the ionizing radiation field,
producing a core of neutral gas. The strongest argument against the HVCs being
very distant has been point d) (see Giovanelli 1977); as a consistency argument
Blitz et al.\ note that Ly$\alpha$-forest absorbers also seem to have small
velocity gradients over large scales.
\par Objection e) was discussed at the workshop. In this context the results of
Zwaan et al.\ (1997) are particularly relevant. They used Arecibo to conduct a
blind \HI\ survey, but found no free-floating starless \HI\ clouds down to a
mass of about 5$\times$10$^7$\,M$_\odot$. Assuming that the Blitz et al.\ model
is correct, one can estimate the mass, distance and radius for each HVC, and the
volume out to which such a cloud could have been detected. Comparing to the
effective detection volume for the Zwaan et al.\ (1997) survey then allows a
prediction for the number of analogous clouds that should have appeared in their
survey. Because of several technical issues, and because a number of free
parameters remain, this is not the appropriate place for a complete discussion
of the resulting predictions. It appears, however, that the Blitz et al.\ model
in its original form is incompatible with the Zwaan et al.\ data; it can be made
compatible by substantially decreasing the relative proportion of \HI\ and thus
substantially decreasing the distance scale of the ensemble. It is as yet
unclear whether these changes will completely exclude a relation between the
HVCs and the Local Group.
\par A crucial test of the Blitz et al.\ model may be provided by the WHAM and
TAURUS-2 Teams, by means of observations of \Ha\ emission of individual clouds.
If these clouds are in the Local Group at large distances, then the incident
radiation field should be weak and the \Ha\ emission becomes undetectable. On
the other hand positive detections of \Ha\ emission for the large majority of
clouds would falsify the Blitz et al.\ model.

\subsection{HVCs as Ly$\alpha$-forest analogues}
\par Mallouris et al.\ (1998 and these proceedings) normalize the distribution
of the number of HVCs per Mpc$^3$ as a function of column density, and find that
both the slope and the absolute normalization fit in the gap in the distribution
between low-redshift Ly$\alpha$ forest clouds (N(\HI)$<$$10^{17}$\,cm$^{-2}$)
and damped Ly$\alpha$ absorbers (N(\HI)$>$$10^{20}$\,cm$^{-2}$). This model
implies that there should not be a turnover in the distribution of column
density, contrary to the interpretation of Blitz et al.\ (1998). The suggestion
is made that the source of the Ly$\alpha$ clouds in general, as well as that of
the HVCs in particular, lies in the tidal disruption of dwarf galaxies. The HVCs
would thus be Local Group objects, though not necessarily as distant as in the
Blitz et al.\ model.

\subsection{The Galactic Fountain}
\par Much progress was also made on the interpretation of HVCs as galactic
objects. De Avillez (this volume) presents the first three-dimensional
hydrodynamic models of the Galactic Fountain. See his Sects.~1.2--1.3 for a
summary of the basics of the Fountain, and for a discussion of the problems with
two-dimensional models. He solves the full equations of motion of the gas in a
gravitational field provided by the stars and dark matter, using the ideal-gas
law for the equation of state and an approximation for the cooling curve.
Supernovae are set off in a manner that is tied to the distribution of
early-type stars, allowing superbubbles to form.
\par The model reproduces many of the general features of the distribution of
cold, cool, warm and hot gas in the Galaxy. A dynamic equilibrium is set up
between upward and downward flowing gas with a rate of
4.2$\times$10$^{-3}$\,M$_\odot$\,kpc$^{-2}$\,yr$^{-1}$, equivalent to
6\,M$_\odot$\,yr$^{-1}$ when integrated over the disk. Chimneys are generated
from superbubbles forming at z$>$100\,pc. The ionized gas forms a layer with a
scaleheight of about 1\,kpc, and is fed by the chimneys and other ascending hot
gas. This warm gas then escapes buoyantly upward, setting up a fountain flow.
I.e., the fountain originates at z=1--1.5\,kpc, rather than in the disk
(z=0\,kpc). From an analytical model, using parameters based on the model, this
fountain gas is predicted to reach z-heights of up to 10\,kpc.
\par Some of the model predictions are relevant to HVC studies. Infalling clouds
form from cooling instabilities in the hot gas in places where shock waves
intersect, creating density variations. The sizes of the cool clouds thus formed
range from a few pc to hundreds of pc in size. The bulk of these clouds have
(infalling) vertical velocities of 20 to 90\,\kms, a small fraction has
v$_z$$<$20\,\kms, and a few have higher velocities (up to $-$300\,\kms). Most of
these clouds occur in a layer between z=1 and 2\,kpc. This height corresponds to
a cool layer that is in unstable equilibrium with the warmer gas below. Clouds
form in complexes with different velocities, and internal dispersions of order
20\,\kms. At higher z a small number of \HI\ clouds are predicted to occur,
because the cooling times get longer. However, the numerical model only extends
up to z=$\pm$4\,kpc, so the number of high-z high-velocity clouds is difficult
to estimate.
\par Thus, it appears that many (or all) of the intermediate-velocity clouds
with $\vert$v$_{\rm LSR}$$\vert$$<$100\,\kms\ can be understood as cool
condensations in the dynamic equilibrium between gravity and supernovae, whereas
this would be the case for only some of the high-velocity clouds. More analysis
of the model results is necessary to determine which fraction of the HVCs can be
understood as part of a fountain flow, and to determine the distribution of LSR
velocities on the sky. Also, it remains to be seen whether the positive
velocities of the clouds at l$>$180\deg\ can be explained as due to projection
effects, or whether these clouds require a separate explanation.

\section{Future prospects}
\par In this final section, we summarize the prospects for future work.
\par From the whole-sky \HI\ surveys (Leiden-Dwingeloo Survey and Villa Elisa
Southern Sky Survey) a more complete catalogue of the HVCs will be made. Most
importantly, this catalogue will also contain the intermediate-velocity clouds.
For the first time such a catalogue will allow detailed studies of cloud
kinematics, linewidths, velocity gradients, as well as a proper statistical
analysis.
\par An important type of study for which little progress was made in the past
few years is to find \HI\ absorption toward background 21-cm continuum sources,
in order to derive spin temperatures. Unlike measurements of \HI\ linewidths,
these provide the kinetic temperatures that are used in model calculations such
as those of Wolfire et al.\ (1995).
\par We foresee much progress in the study of ionized hydrogen at high velocity,
using WHAM. Of particular interest are the following questions: a) is
photo-ionization the main source of ionization? b) can we estimate cloud
distances by combining the \Ha\ intensity with a model of the galactic radiation
field? c) does the ionized hydrogen form a shell around a neutral core, or is it
pervasive? d) how important is H$^+$ in deriving abundances? By combining
observations of optical emission lines with UV absorption lines it will also
become possible to measure the temperature and density of the ionized gas, as
well as constrain the cloud structure. Combining [NII]/\Ha\ with [SII]/\Ha\
(both dependent on abundance and temperature) will allow to set constraints on
abundances. The \Ha\ measurements alone will give strong constraints on the
possibility that the HVCs are Local Group objects.
\par Further progress on metal abundances is expected with FUSE, especially for
complex~C, the Magellanic Stream and HVC287+22+240, and possibly for complexes~A
and M. FUSE will also provide much data on depleted elements, and thus allow a
comparison of the depletion pattern with other kinds of gas (such as cool disk
gas, warm disk gas and halo gas). FUSE data on high-velocity \OVI\ will provide
a better understanding of the hot gaseous halo of the Galaxy. More HST data for
\SII\ in different HVCs and IVCs could provide better insight into whether
clouds have a galactic (Fountain) or an extra-galactic (Accretion) origin.
\par Work on distance determinations should profit from a more organized
approach to finding and observing suitable background probes. While for several
large HVCs significant lower distance limits are available or may be readily
obtained, special probe searches will be required in order to establish upper
distance limits, and also to obtain constraints on the distances of smaller
HVCs.
\par The recognition of the leading counterpart to the trailing Magellanic
Stream has provided the possibility of improved dynamical work on the Magellanic
System. The positions and velocities of these leading-bridge clouds give extra
constraints on the orbits of the Magellanic Clouds. The addition of gas dynamics
to n-body models gives the possibility of studying the structure of the outer
Galactic Halo. The fact that small-scale structure is present in the Stream in a
probable outer-Halo environment should provide constraints on models for the
physical conditions in that Halo.
\par The study of the relation to the underlying galaxy of high-velocity cloud
analogues in nearby spiral galaxies (such as M\,31, M\,33, M\,51, M\,83, M\,101,
NGC\,628, NGC\,891, NGC\,1300, NGC\,5668) can provide more information about the
relative importance of the Fountain vs Accretion.
\par In summary, in the next few years we expect to see substantial progress in
measuring the properties of high-velocity gas, such as distances, metallicities,
depletion patterns, and kinematics. Dynamical models of the Galactic Fountain,
Magellanic Stream and Accretion will allow progress on understanding the origins
of HVCs.


\begin{references}
\def\ref{\reference}
\ref Arnal M., Bajaja E., Morras R., P\"oppel W., 1996, RMxAC, 4, 132
\ref Bates B., Catney M.G., Keenan F.P., 1990, \mnras, 242, 267
\ref Benjamin R.A., Danly L., 1997, \apj, 481, 764
\ref Bland-Hawthorn J. Maloney P.R. 1998, \apj, in press (astro-ph/9810469)
\ref Bland-Hawthorn J., Veillieux S., Cecil G.N., Putman M.E., Gibson B.K.,
     Maloney P.R., 1998, \mnras, 299, 611
\ref Blitz L., Spergel D., Teuben P., Hartmann D., Burton W.B., 1998, \apj, in
     press (astro-ph/9803251)
\ref Blom J.J., Bloemen H., Bykov A.M., Burton W.B., Hartmann D., Hermsen W.,
     Iyudin A.F., Ryan J., Schoenfelder V., Strong A.W., Uvarov Y., 1997, \aap,
     321, 288
\ref Brown P.J.F., Dufton P.L., Keenan F.P., Boksenberg A., King D.L., Pettini
     M., 1989, \apj, 339, 397
\ref Burrows D.N., Mendenhall J.A., 1991, Nature, 351, 629
\ref Christodoulou D.M., Tohline J.E., Keenan F.P., 1997, \apj, 486, 810
\ref Comeron F., Torra J., 1992, \aap, 261, 94
\ref Comeron F., Torra J., 1994, \aap, 281, 35
\ref Cox D.P., Reynolds R.J., 1987, \araa, 25, 303
\ref Danly L., Albert C.E., Kuntz K.D., 1993, \apj, 416, L29
\ref Falgarone E., Phillips T.G., Walker C.K., 1991, \apj, 378, 186
\ref Gardiner L.T., Noguchi M., 1996, \mnras, 278, 191
\ref Giovanelli R., 1977, \aap, 55, 395
\ref Giovanelli R., 1980, \aj, 85, 1155
\ref Gladders M.D., Clarke T.E., Burns C.R., Attard A., Casey M.P., Hamilton D.,
     Mallin-Ornelas G.,Karr J.L., Poirier S.M., Sawicki M., Barrientos L.F.,
     Mochnacki S.W., 1998, \apj, 507, L161
\ref Hartmann D., Burton W.B., 1997, Atlas of Galactic Neutral Hydrogen,
     (Cambridge University Press)
\ref Herbstmeier U., Kalberla P.M.W., Mebold U., Weiland H., Souvatzis I.,
     Wennmacher A., Schmitz J., 1996, \aaps, 117, 497
\ref Hulsbosch A.N.M., Wakker B.P., 1988 \aaps, 75, 191
\ref Ivezi\'c Z., Christodoulou D.M., 1997, \apj, 486, 818
\ref Keenan F.P., Shaw C.R., Bates B., Dufton P.L., Kemp S.N., 1995, \mnras,
     272, 599
\ref Kerp J., Burton W.B., Egger R., Freyberg M.J., Hartmann D.,
     Kalberla P.M.W., Mebold U., Pietz J., \aap, 1998, in press
     (astro-ph/9810307)
\ref Kulkarni S.R., Dickey J.M., Heiles C., 1985, \apj, 291, 716
\ref Lepine J.R.D., Duvert G., 1994, \aap, 286, 60
\ref Lockman F.J., Jehoda K., McCammon D., 1986, \apj, 302, 432
\ref Lu L., Savage B.D., Sembach K.R., 1994, \apj, 426, 563
\ref Lu L., Savage B., Sembach K., Wakker B., Sargent W.., Oosterloo T., 1998,
     \aj, 115, 162
\ref Majewski S.R., Ostheimer J.C., Kunkel W.E., Johnston K.V. Patterson R.J.,
     1999, in New Views of the Magellanic Clouds, ASP Conf.\ Series, in press
\ref Mallouris C., York D.G., Lanzetta K., 1998, \baas, 192, \#51.09
\ref Mathewson D.S., Cleary M.N., Murray J.D., 1974, \apj, 190, 291
\ref McCammon D., Sanders W.T., 1990, \araa, 28, 657
\ref Moore B., Davis M., 1994, \mnras, 270, 209
\ref Morras R., Bajaja E., Arnal E.M., 1998, \aap, 334, 659
\ref Murphy E.M., Lockman F.J., Savage B.D., 1995, \apj, 447, 642
\ref Oort J.H., 1970, \aap, 7, 381
\ref Oosterloo T., Da Silva C.G., Staveley-Smith L., 1996, \aj, 112, 1969
\ref Ostheimer J.C., Majewski S.R., Kunkel W.E., Johnston K.V., 1997, \baas,
     191, \#131.03
\ref Putman M.E., Gibson B.K., Staveley-Smith L., et al., 1998, Nature, 394, 752
\ref Reynolds R.J., 1985, ApJ, 294, 256
\ref Ryans R.S.I., Keenan F.P., Sembach K.R., Davies R.D., 1997a, \mnras, 289,
     83
\ref Ryans R.S.I., Keenan F.P., Sembach K.R., Davies R.D., 1997b, \mnras, 289,
     986
\ref Sahu M.S., 1998, \aj, 116, 1205
\ref Sahu M.S., Blades J.C., 1997, \apj, 484, L125
\ref Schulman E., Bregman J.N., Brinks E., Roberts M.S., 1996, \aj, 112, 960
\ref Schulman E., Brinks E., Bregman J.N., Roberts M.S., 1997a, \aj, 113, 1559
\ref Schulman E., Ockels F., Knezek P.M., 1997b, \baas, 191, \#77.06
\ref Schwarz U.J., Wakker B.P., van Woerden H., 1995, \aap, 302, 364
\ref Sembach K.R., Savage B.D., Lu L., Murphy E.M., ApJ, 1998, in press
     (astro-ph/9809231)
\ref Staveley-Smith L., 1997, PASA 14, 111
\ref Stoppelenburg P.S., Schwarz U.J., van Woerden H., 1998, \aap, 338, 200
\ref Swaters R.A., Sancisi R., van der Hulst J.M., 1997, \apj, 491, 150
\ref Tamanaha C.M., 1995, \apj, 450, 638
\ref Tamanaha C.M., 1996, \apjs, 104, 81
\ref Tamanaha C.M., 1997, \apjs, 109, 139
\ref Tufte S.L., Reynolds R.J., Haffner L.M., 1998, \apj, 504, 773
\ref Vogelaar M.G.R., Wakker B.P., 1994, \aap, 291, 557
\ref Wakker B.P., Schwarz U.J., 1991, \aap, 250, 484
\ref Wakker B.P., Vijfschaft B., Schwarz U.J., 1991, \aap, 249, 233
\ref Wakker B.P., van Woerden H., 1991, \aap, 250, 509
\ref Wakker B.P., van Woerden H., 1997, \araa, 35, 217
\ref Wakker B.P., Howk J.C., Savage B.D., Tufte S.L., Reynolds R.J., van
     Woerden H., Schwarz U.J., Peletier R.F., 1999, these proceedings
\ref Wakker B.P., Savage B.D., Oosterloo T.A., Putman M.E., 1999, these
     proceedings
\ref Weiner B.J., Williams T.B., 1996, \aj, 111, 1156
\ref Wisotzki L., Koehler T., Groote D., Reimers D., 1996, \aaps, 115, 227
\ref van Woerden H., Schwarz U.J., Peletier R.F., Wakker B.P., Kalberla P.M.W.,
     1999, Nature, submitted
\ref Wolfire M.G., McKee C.F., Hollenbach D., Tielens A.G.G.M., 1995, \apj, 453,
     673
\ref Yoshizawa A. 1998, PhD Thesis, Tohoku University, Sendai, Japan
\ref Zimmer F., Lesch H., Birk G.T., 1997, \aap, 320, 746
\ref Zwaan M.A., Briggs F.H., Sprayberry D., 1997, \apj, 490, 173
\end{references}
\end{document}